\documentclass[pra,aps,amsfonts,amsmath,twocolumn,showpacs,floatfix]{revtex4}
\usepackage{bm}
\usepackage{amsfonts}
\usepackage{amsmath}
\usepackage{graphicx}
\newcommand{\ev}{\bm{\mathrm{e}}}
\newcommand{\jv}{\bm{\mathrm{j}}}
\newcommand{\pv}{\bm{\mathrm{p}}}
\newcommand{\rv}{\bm{\mathrm{r}}}

\newcommand{\nablav}{\bm{\nabla}}
\newcommand{\Omegav}{\bm{\Omega}}
\newcommand{\Deltatil}{\tilde{\Delta}}
\newcommand{\htil}{\tilde{h}}
\newcommand{\kappatil}{\tilde{\kappa}}
\newcommand{\rhotil}{\tilde{\rho}}
\newcommand{\Deltab}{\bar{\Delta}}
\newcommand{\rhob}{\bar{\rho}}
\newcommand{\omegab}{\bar{\omega}}
\newcommand{\poisson}[2]{\{#1,#2\}}
\newcommand{\bigpoisson}[2]{\big\{#1,#2\big\}}
\newcommand{\Bigpoisson}[2]{\Big\{#1,#2\Big\}}
\newcommand{\Eq}[1]{Eq.\@ (\ref{#1})}
\newcommand{\Eqs}[1]{Eqs.\@ (\ref{#1})}
\newcommand{\Ref}[1]{Ref.\@ \cite{#1}}
\newcommand{\Fig}[1]{Fig.\@ \ref{#1}}
\newcommand{\Figs}[1]{Figs.\@ \ref{#1}}
\begin{document}
\title{Two-fluid model for a rotating trapped Fermi gas in the BCS phase}
\author{Michael Urban}
\affiliation{Institut de Physique Nucl{\'e}aire, F-91406 Orsay
C{\'e}dex, France}
\begin{abstract}
We investigate the dynamical properties of a superfluid gas of trapped
fermionic atoms in the BCS phase. As a simple example we consider the
reaction of the gas to a slow rotation of the trap. It is shown that
the currents generated by the rotation can be understood within a
two-fluid model similar to the one used in the theory of
superconductors, but with a position dependent ratio of normal and
superfluid densities. The rather general result of this paper is that
already at very low temperatures, far below the critical one, an
important normal-fluid component appears in the outer regions of the
gas. This renders the experimental observation of superfluidity
effects more difficult and indicates that reliable theoretical
predictions concerning other dynamical properties, like the
frequencies of collective modes, can only be made by taking into
account temperature effects.
\end{abstract}
\maketitle
In the last few months, experiments with trapped fermionic $^6$Li
atoms made great progress. The fact that by using the Feshbach
resonance the Fermi gas can be transformed into a Bose-Einstein
condensate (BEC) of molecules, which can be cooled by evaporative
cooling and afterwards transformed back into a Fermi gas, allows to
reach extremely low temperatures of the order of $0.03 T_F$
\cite{Bartenstein}, where $T_F = k_B \epsilon_F$ is the Fermi
temperature. This allows, among other things, a detailed study of the
BEC-BCS crossover. In particular, a temperature of $0.03 T_F$ should
even be low enough to realize the BCS phase which is characterized by
the condition $\Delta\ll\epsilon_F$, where $\Delta$ denotes the
pairing gap.

However, at present it is not very clear how the transition to the BCS
phase could be detected. While several observables related to
collective oscillations (e.g. breathing modes) of the system have been
investigated \cite{Bartenstein,Kinast}, the most unambigious
signatures of the superfluid BCS phase seem to be those which concern
the rotational properties of the system \cite{Cozzini}. For instance,
the moment of inertia of a slowly rotating Fermi gas was proposed to
be a suitable observable for the detection of the BCS transition
\cite{Farine}. It should be mentioned that at present most of the
theoretical predictions concerning possible experimental signatures of
the BCS phase (e.g., \cite{Minguzzi,Zambelli,Cozzini,Menotti,Stringari})
neglect temperature effects as well as possible deviations from
hydrodynamic behaviour due to the discrete level spectrum in the trap.

In a previous article \cite{UrbanSchuck} we calculated the moment of
inertia of a superfluid atomic Fermi gas in a slowly rotating trap at
finite temperature. There it turned out that the irrotational flow,
which is characteristic for superfluidity, is realized only in the
limit when the gap $\Delta$ is very large compared with the
temperature $T$ and the level spacing $\hbar \omega$ of the trap. In
all other cases, the velocity field has both, rotational and
irrotational components. For example, if the level spacing $\hbar
\omega$ is comparable with $\Delta$, the current has a strong
rotational component even at zero temperature. On the other hand, at
nonvanishing temperature $T$, a certain fraction of the Cooper pairs
is broken by thermal excitations. This leads to the well-known effect
that the system behaves like a mixture of normal and superfluid
components \cite{Leggett,BetbederMatibet,Schrieffer}. Under rotation,
the former behaves like a rigid body, while the latter can only have
an irrotational velocity field.

However, in the calculation of \Ref{UrbanSchuck} the gap $\Delta(\rv)$
has been replaced by a constant $\Delta$ corresponding to the average
diagonal matrix element of $\Delta(\rv)$ at the Fermi surface. While
this averaging procedure seems to be justified in cases where only one
oscillator shell participates in the pairing (intrashell pairing,
$\Delta<\hbar \omega$), it is not well suited for the strong pairing
regime ($\Delta > \hbar\omega$), where the properties of the system
can be described locally and depend on $\rv$ via the spatial
dependence of $\Delta(\rv)$ \cite{Heiselberg}. In particular, the
normal and superfluid fractions of the density, $\rho_n/\rho$ and
$\rho_s/\rho$, should depend on $\rv$. To our knowledge this fact has
not been taken into account in the existing literature.

In this article, we will concentrate on the $\hbar\to 0$ limit, i.e.,
we will neglect the quantum effect which is responsable for the
rotational component of the velocity field at zero temperature. Anyway,
if the system is sufficiently large and if the temperature is not
extremely low, this quantum effect becomes much smaller than the
effect resulting from the thermally created normal component of the
system. The important point is that we will now take into account the
$\rv$ dependence of the gap. In addition, we will not rely on the
simplification made in our previous work that the full potential (trap
$+$ mean field) is approximately harmonic.

Let us briefly summarize the most important formulas (for more
explanations and details see \Ref{UrbanSchuck}). We assume that equal
numbers of atoms in two spin states are trapped in a harmonic
potential,
\begin{equation}
V_\mathit{trap}(\rv) =
\sum_{i=xyz}\frac{m\omega_i^2}{2}r_i^2\,.
\label{vtrap}
\end{equation}
The cigar-shaped form of the traps used in current experiments
corresponds to $\omega_z \ll \omega_x = \omega_y$. However, in order
to force the system to rotate around the long axis, one has to
break the axial symmetry, e.g., by using a rotating laser beam as
``spoon''. We will model this by taking $\omega_x\neq\omega_y$. The
mean-field single-particle hamiltonian minus the chemical potential
reads
\begin{equation}
\hat{h}_0 = \frac{\hat{\pv}^2}{2m} + V_\mathit{trap}(\rv)+g\, \rho(\rv)
-\mu\,,
\label{hsp}
\end{equation}
the coupling constant $g = 4\pi\hbar^2 a/m$ being proportional to the
atom-atom scattering length $a < 0$, and $\rho(\rv)$ being the density
per spin state. The order parameter in equilibrium is denoted
$\Delta_0(\rv)$.

Now we want to describe what happens if the trap is slowly rotating
around the $z$ axis with a rotation frequency $\Omegav = \Omega
\ev_z$. This is most easily done in the rotating reference frame. Then
we still have a static problem, but the hamiltonian receives the
additional term
\begin{equation}
\hat{h}_1 = -\Omega \hat{L}_z = -(\Omegav\times\rv)\cdot\hat{\pv}\,,
\end{equation}
which we will treat as a small perturbation. Because of this term, the
order parameter $\Delta(\rv)$ receives a phase
$\exp[-2i\phi(\rv)]$. The explicit form of $\phi(\rv)$ is unknown for
the moment and will be determined below. It is convenient to eliminate
this phase by a gauge transformation, multiplying all single-particle
wave functions by $\exp[i\phi(\rv)]$. In this way the gauge
transformed gap $\Deltatil(\rv)$ stays real, which in the case of a
slow rotation implies that $\Deltatil$ does not change at all
\footnote{Since the magnitude of $\Delta$ cannot depend on the sign of
$\Omega$, its change must be at least of the order $\Omega^2$.},
i.e., $\Deltatil = \Delta_0$. On the other hand, this gauge
transformation changes the momentum operator according to
\begin{equation}
\hat{\tilde{\pv}} = \hat{\pv}-\hbar\nablav\phi(\rv)\,.
\end{equation}
Hence, the price to pay for the real gap is an additional term in the
hamiltonian. To linear order in the rotation frequency the new
perturbation hamiltonian reads
\begin{equation}
\hat{\htil}_1 = -\Omega \hat{L}_z-\frac{\hbar}{2m}
  \big(\hat{\pv}\cdot[\nablav\phi(\rv)]
  +[\nablav\phi(\rv)]\cdot\hat{\pv}\big)\,.
\label{h1tilde}
\end{equation}

In order to describe the system semiclassically, we make use of the
Wigner-Kirkwood expansion. To that end we denote the Wigner transforms
of $\hat{h}_0$, $\hat{\htil}_1$, etc., by $h_0(\rv,\pv)$,
$\htil_1(\rv,\pv)$, etc. We need also the Wigner transforms of the
normal and abnormal density matrices in equilibrium, $\rho_0(\rv,\pv)$
and $\kappa_0(\rv,\pv)$, as well as their deviations from equilibrium,
$\rhotil_1(\rv,\pv)$ and $\kappatil_1(\rv,\pv)$. (For the sake of
brevity, we will occasionally omit the arguments $\rv$ and $\pv$ if
there is no risk of confusion.) Furthermore we introduce the Poisson
bracket of two phase-space functions
\begin{equation}
\poisson{f}{g} = \sum_{i=x,y,z}
 \Big(\frac{\partial f}{\partial r_i}\,\frac{\partial g}{\partial p_i}
 -\frac{\partial f}{\partial p_i}\,\frac{\partial g}{\partial r_i}\Big)\,.
\end{equation}
Using the notations defined above, the terms linear in $\Omega$ of the
Hartree-Fock-Bogoliubov (HFB) or Bogoliubov-de Gennes equations up to
linear order in $\hbar$ can be written as
\begin{gather}
i\hbar\poisson{h_0}{\rhotil_1}+2\Delta_0 \kappatil_1 =
  -i\hbar\bigpoisson{\htil_1}{\rho_0}\,\label{hfb1}\\
i\hbar\poisson{\Delta_0}{\rhotil_1}-2h_0 \kappatil_1 =
  i\hbar\bigpoisson{\htil_1}{\kappatil_0}\,.\label{hfb2}
\end{gather}
These are exactly the Eqs.\@ (84) and (85) of \Ref{UrbanSchuck}. The
main point of the present article concerns the solution of this system
of equations in the case of an $\rv$-dependent gap $\Delta_0(\rv)$.

First we eliminate $\kappatil_1$ by multiplying \Eq{hfb1} by $h_0$ and
\Eq{hfb2} by $\Delta_0$ and adding up the two resulting
equations. Using the chain and product rules of differentiation, we
then obtain
\begin{equation}
\frac{1}{2}\bigpoisson{E^2}{\rhotil_1} =
  -h_0\bigpoisson{\htil_1}{\rho_0}
  +\Delta_0\bigpoisson{\htil_1}{\kappa_0}\,,
\label{hfb4}
\end{equation}
where $E^2(\rv,\pv) = h_0^2(\rv,\pv) + \Delta_0^2(\rv)$. To proceed
further, we express $\rho_0$ and $\kappa_0$ in terms of $h_0$ and
$\Delta_0$. Within the Thomas-Fermi (TF) or local-density
approximation, these relations read
\begin{gather}
\rho_0(\rv,\pv) =
\frac{1}{2}-\frac{h_0(\rv,\pv)}{2E(\rv,\pv)}\big(1-2f[E(\rv,\pv)]\big)\,,
\label{rhotf}\\
\kappa_0(\rv,\pv) =
\frac{\Delta_0(\rv)}{2E(\rv,\pv)}\big(1-2f[E(\rv,\pv)]\big)\,,
\label{kappatf}
\end{gather}
where $f(E) = 1/[\exp(E/k_BT)+1]$ denotes the Fermi function. Although
\Eqs{rhotf} and (\ref{kappatf}) are the solutions of the $\hbar\to 0$
limit of the HFB equations, they are valid up to linear order in
$\hbar$ \cite{Taruishi} and therefore consistent with \Eqs{hfb1} and
(\ref{hfb2}). Inserting \Eqs{rhotf} and (\ref{kappatf}) into
\Eq{hfb4}, we obtain, again after repeated use of chain and product
rules of differentiation, the following simple equation:
\begin{equation}
\frac{1}{2}\bigpoisson{E^2}{\rhotil_1} =
  \frac{1}{2}\Bigpoisson{E^2}{\htil_1\frac{df}{dE}}\,.
\end{equation}
It is evident that this equation is solved by
\begin{equation}
\rhotil_1 = \htil_1\frac{df}{dE}\,.
\label{solutionhfb}
\end{equation}

However, before this solution can be used, the gauge transformation
which has been introduced in order to make the gap real must be
inverted:
\begin{equation}
\rho(\rv,\pv) = \rhotil[\rv,\pv+\hbar\nablav\phi(\rv)]\,.
\end{equation}
To linear order in $\Omega$, this can also be written as
\begin{align}
\rho(\rv,\pv) =& \rho_0[\rv,\pv+\hbar\nablav\phi(\rv)] +
    \rhotil_1(\rv,\pv)\\
  =& \rho_0[\rv,\pv+\hbar\nablav\phi(\rv)]\nonumber \\
   &-\Big(\Omegav\times\rv+\frac{\hbar}{m}\nablav\phi(\rv)\Big)\cdot \pv
  \frac{df}{dE}\Big|_{E(\rv,\pv)}\,.
\end{align}
The last line has been obtained with the help of \Eqs{h1tilde} and
(\ref{solutionhfb}).

\begin{figure}
\includegraphics[scale=1.4]{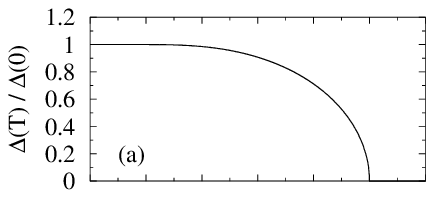}\\[-3mm]
\includegraphics[scale=1.4]{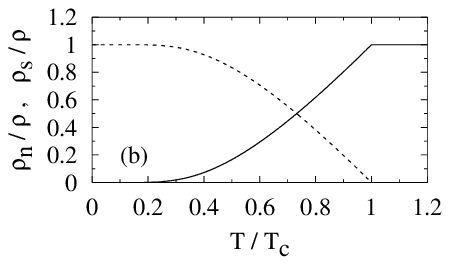}
\caption{\label{Fig1} (a) Temperature dependence of the gap normalized
to its value at zero temperature ($\Delta(0) = 1.76 T_c$). (b)
Temperature dependence of the normal (solid line) and superfluid
(deshed line) fractions of the superfluid system in the limit of weak
pairing ($\Delta \ll \epsilon_F$). These relations hold locally, if
$T_c$ is interpreted as the local critical temperature, defined by
$T_c(\rv) = 0.57 \Delta(\rv;T=0)$.}
\end{figure}

The next step consists in calculating the corresponding current
density per spin state,
\begin{equation}
\jv(\rv) = \int\!\frac{d^3p}{(2\pi\hbar)^3}\,\frac{\pv}{m}\,\rho(\rv,\pv)\,.
\end{equation}
Using the explicit expression for $\rho(\rv,\pv)$ given above, one easily
obtains
\begin{align}
\jv(\rv) = &-\rho_0(\rv) \frac{\hbar}{m}\nablav\phi(\rv)\\
  &-\Big(\Omegav\times\rv+\frac{\hbar}{m}\nablav\phi(\rv)\Big)
    \int_0^\infty \frac{dp}{6\pi^2\hbar^3} p^4\frac{df}{dE}\Big|_{E(\rv,p)}\,,
\end{align}
with
\begin{equation}
\rho_0(\rv) = \int\!\frac{d^3p}{(2\pi\hbar)^3} \rho_0(\rv,\pv)\,.
\end{equation}
The current density can therefore be written in a more suggestive way
in terms of normal and superfluid densities $\rho_n(\rv)$ and
$\rho_s(\rv)$,
\begin{equation}
\jv(\rv) = \rho_n(\rv) \Omegav\times\rv
  -\rho_s(\rv) \frac{\hbar}{m}\nablav\phi(\rv)\,,
\label{current}
\end{equation}
if the normal and superfluid densities are defined according to the
textbook result (\Ref{FetterWalecka}, p.\@ 459) as
\begin{equation}
\rho_n(\rv) = \rho_0(\rv)-\rho_s(\rv)
  = -\int_0^\infty \frac{dp}{6\pi^2\hbar^3}
    p^4\frac{df}{dE}\Big|_{E(\rv,p)}\,.
\label{rhon}
\end{equation}
In the BCS limit, i.e., if $\Delta_0(\rv) \ll \epsilon_F(\rv)$, where
$\epsilon_F(\rv) = \mu-V_\mathit{trap}(\rv)-g\rho_0(\rv)$ denotes the
local Fermi energy, the ratio $\rho_n(\rv)/\rho_0(\rv)$ becomes a
function of only one dimensionless argument, $T/T_c(\rv)$, where
$T_c(\rv) = 0.57 \Delta_0(\rv;T=0)$ denotes the local critical
temperature (the existence of a local critical temperature is an
artifact of the TF approximation). This function, as well as the
temperature dependence of the ratio $\Delta_0(\rv)/\Delta_0(\rv;T=0)$,
is shown in \Fig{Fig1}.

Up to now, the phase $\phi(\rv)$, which determines the velocity of the
superfluid component, is completely unknown. In \Ref{UrbanSchuck},
this phase was determined by calculating $\Deltatil_1$ from
$\kappatil_1$ and imposing the condition $\Deltatil_1=0$. Then it was
shown that with this choice the continuity equation for the current
was satisfied. Here we will adopt another method which is commonly
used in the literature \cite{BetbederMatibet} and which consists in
using the continuity equation. In the rotating frame the latter reads
\begin{equation}
\nablav\cdot\jv(\rv) + \dot{\rho}(\rv)
  - (\Omegav\times\rv)\cdot\nablav\rho(\rv) = 0\,,
\end{equation}
where $\dot{\rho}(\rv)=0$ in our case of a stationary rotation and
$\rho(\rv) = \rho_0(\rv)$ up to linear order in $\Omega$. Taking the
divergence of \Eq{current}, one can see that the normal fluid
component drops out and it remains a continuity equation for the
superfluid component:
\begin{equation}
-\frac{\hbar}{m}\nablav\cdot
  [\rho_s(\rv)\nablav\phi(\rv)]
 -(\Omegav\times\rv)\cdot\nablav\rho_s(\rv) = 0\,.
\label{continuity}
\end{equation}

In the case of a deformed harmonic trapping potential, this equation
can be solved analytically. To see this, remember that within the TF
approximation the density $\rho_0(\rv)$ and the gap $\Delta_0(\rv)$,
and consequently also the superfluid density $\rho_s$, depend on $\rv$
only via the local chemical potential $\mu_\mathit{loc}(\rv) = \mu -
V_\mathit{trap}(\rv)$, i.e., $\rho_s(\rv) = \rho_s[\mu -
V_\mathit{trap}(\rv)]$. Hence \Eq{continuity} can be written as
\begin{multline}
\frac{d\rho_s}{d\mu_\mathit{loc}}\Big|_{\mu-V_\mathit{trap}(\rv)}
  [\nablav V_\mathit{trap}(\rv)]\cdot
  \Big(\frac{\hbar}{m}\nablav\phi(\rv)+\Omegav\times\rv\Big)\\
  -\frac{\hbar}{m}\rho_s(\rv)\nablav^2\phi(\rv) = 0\,.
\end{multline}
In the special case of the harmonic potential (\ref{vtrap}), it can
readily be shown that this equation has the same solution as in the
simple case of constant $\Delta_0$ studied in \Ref{UrbanSchuck},
\begin{equation}
\phi(\rv) = \frac{m}{\hbar}\,
  \frac{\omega_x^2-\omega_y^2}{\omega_x^2+\omega_y^2}\,\Omega r_x
  r_y\,,
\end{equation}
since this solution is independent of the form of $\rho_s(\mu)$. The
current density per spin state, $\jv(\rv)$, is therefore given by
\begin{equation}
\jv(\rv) = \rho_n(\rv)\Omegav\times\rv-\rho_s(\rv)
  \frac{\omega_x^2-\omega_y^2}{\omega_x^2+\omega_y^2}\,\Omega
  \nablav(r_x r_y)\,.
\label{currentexplicit}
\end{equation}

Due to fact that $\rho_0(\rv)$ and $\Delta_0(\rv)$ [and consequently
$\rho_s(\rv)$ and $\rho_n(\rv)$] depend on $\rv$ only via the local
chemical potential, it is sufficient to perform the TF calculation for
a spherical trap with the geometrically averaged trapping frequency
$\bar{\omega} = (\omega_x\omega_y\omega_z)^{1/3}$. In this spherical
trap, of course, the density $\rhob_0$, gap $\bar{\Delta}_0$, etc.,
depend only on the distance from the center, i.e., $\rhob_0(\rv)
= \rhob_0(r)$, etc. (quantities related to the spherical trap will be
marked by a bar). The corresponding quantities in the deformed trap
can then be obtained from
\begin{equation}
\rho_0(\rv) = \rhob_0\Big(\frac{1}{\bar{\omega}}
  \sqrt{\omega_x^2 r_x^2+\omega_y^2 r_y^2+\omega_z^2 r_z^2}\Big)\,
\label{deformed}
\end{equation}
and analogously for $\Delta_0(\rv)$, $\rho_s(\rv)$, etc. Note,
however, that all this is true only within the TF approximation.

\begin{figure}
\includegraphics[scale=1.4]{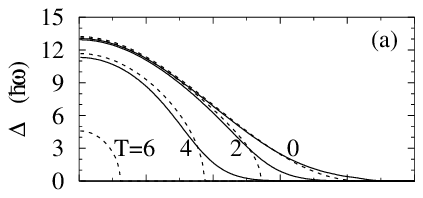}\\[-3mm]
\includegraphics[scale=1.4]{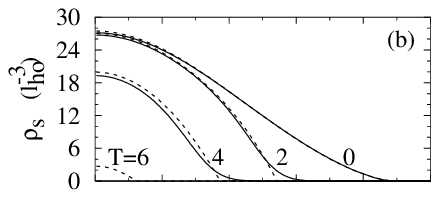}\\[-3mm]
\includegraphics[scale=1.4]{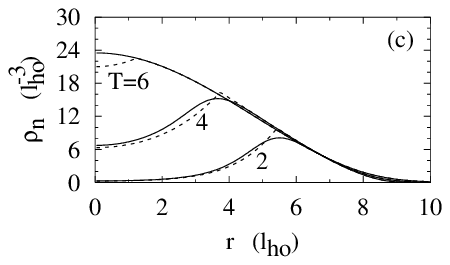}
\caption{\label{Fig2} (a) gap $\Deltab_0(r)$, (b) superfluid density
$\rhob_s(r)$, and (c) normal-fluid density $\rhob_n(r)$ as a function
of the distance $r$ from a spherical trap with frequency $\omegab$ for
several temperatures: $T = 0,2,4,6\, \hbar\omegab/k_B$. For
convenience, all quantities are given in harmonic oscillator units,
i.e., $r$ in $l_{ho} = \sqrt{\hbar/m\omegab}$, $\Deltab_0$ in
$\hbar\omegab$, $\rhob_s$ and $\rhob_n$ in $l_{ho}^{-3}$. The number
of atoms is 36000, their ineraction strength is set to $g = -\hbar^2
l_{ho}/m$. This choice of parameters allows us to compare the TF
results (dashed lines) with results obtained from a numerical solution
of the full HFB equations (solid lines). Note, however, that within
the TF approximation one obtains qualitatively similar results in the
case of more realistic parameters, if the temperatures are scaled with
the corresponding critical temperature.}
\end{figure}

In \Fig{Fig2} we show the normal and superfluid densities per
spin state ($\rhob_n$ and $\rhob_s$) and the gap $\Deltab_0$ in the
spherical trap for different temperatures. The dashed lines correspond
to the TF ($\hbar\to 0$) results, while for the solid lines the gap
$\Deltab_0$ has been obtained by solving numerically the HFB equations
\cite{GrassoUrban}. At zero temperature, the trapped Fermi gas is
completely superfluid, i.e., $\rhob_s = \rhob_0$ and $\rhob_n = 0$. At
low but non-vanishing temperature ($T = 2 \hbar\omegab \approx 0.36
T_c$, $T_c\approx 5.5\hbar\omegab$ being defined as the critical
temperature within HFB), a normal fluid component appears near the
surface, since there the gap is smallest and consequently the Cooper
pairs are most easily broken by thermal excitations. If the
temperature increases further ($T = 4\hbar\omegab \approx 0.72 T_c$),
the normal-fluid component starts to extend over the whole volume, and
finally, slightly above the critical temperature ($T = 6\hbar\omegab$)
the superfluid component vanishes completely (solid lines). The small
superfluid region near the center of the trap which survives at this
temperature within the TF approximation (dashed line) is a consequence
of the fact that the critical temperature predicted by the TF
approximation is higher than the HFB one
\cite{BaranovPetrov1}. However, apart from this point, one can say that
in general for $\rhob_n$ and $\rhob_s$ the agreement between TF and
HFB is reasonable and better than for the gap $\Deltab_0$ itself. The
reason why the agreement between TF and HFB is better for $\rhob_n$
and $\rhob_s$ than for $\Deltab_0$ is that near the critical
temperature the temperature dependence of $\rho_s/\rho$ is much weaker
than that of $\Delta/\Delta(T=0)$, cf.\@ \Fig{Fig1}.

In order to make the comparison between the gaps $\Deltab_0$
calculated within the TF approximation and by solving the full HFB
equations, we had to choose a rather moderate number of particles, $N
= 36000$, for which the HFB calculation is feasible. If we had not
been interested in the comparison between the results obtained with
the HFB and TF gaps, we could of course have shown the TF results for
arbitrarily large numbers of particles. We emphasize that even for
much larger numbers of particles the qualitative behaviour of the TF
results remains unchanged, provided that the coupling constant $g$ is
tuned such that the condition $\Delta \ll \epsilon_F$ (BCS condition)
remains satisfied
\footnote{At zero temperature and at the center of the trap we have
with our choice of parameters $\Delta\approx 0.2 \epsilon_F$, i.e., we
are already close to the BCS-BEC crossover region.}
and the temperatures are scaled with respect to the critical
temperature.

\begin{figure}
\includegraphics[scale=1.4]{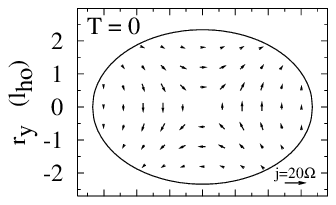}\hspace{-3mm}
\includegraphics[scale=1.4]{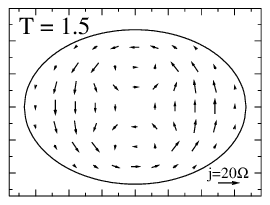}\\[-4mm]
\includegraphics[scale=1.4]{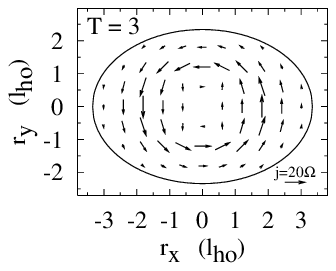}\hspace{-3mm}
\includegraphics[scale=1.4]{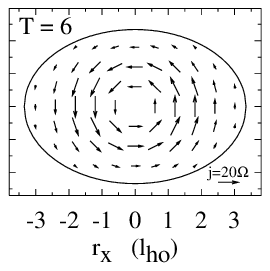}
\caption{\label{Fig3} Current density divided by the angular velocity
of the rotation, $\jv(\rv)/\Omega$, in the $xy$ plane ($r_z = 0$) of a
trap with $\omega_x/\omega_y = 0.7$ and $\omega_z/\omega_r = 0.03$
($\omega_r = \sqrt{\omega_x\omega_y}$) at four different
temperatures. The remaining parameters are the same as in
\Fig{Fig2}.The length of the arrow displayed in the lower right corner
of each figure corresponds to a current density of $j = 20
\Omega/l_{ho}^2$. The coordinates $r_x$ and $r_y$ are given in units
of $l_{ho}$ and the temperatures in units of $\hbar\omegab/k_B$.}
\end{figure}

Using the spherical density profiles and \Eqs{deformed} and
(\ref{currentexplicit}), we can immediately calculate the current
distribution $\jv(\rv)$. In \Fig{Fig3} we show the current in the $xy$
plane ($r_z = 0$) for a deviation from axial symmetry of
$\omega_x/\omega_y = 0.7$ at several temperatures. (For the cases
shown, the current densities obtained from the HFB and TF density
profiles are indistinguishable within the resolution of the plot.) At
zero temperature, the current is irrotational and rather weak (it
vanishes in the limit of axial symmetry, $\omega_x =\omega_y$). In the
surface region the current reaches its ordinary (rigid-body) form
already at $T = 1.5 \hbar\omegab\approx 0.27 T_c$). At $T =
3\hbar\omegab\approx 0.54 T_c$ the current shows almost everywhere the
rigid-body behavior, only near the center it is still a little bit
weaker than in the normal phase, $T = 6\hbar\omegab$.

Let us now look at the temperature dependence of the moment of
inertia, $\Theta$, which is defined as
\begin{equation}
\Theta = \frac{\langle \hat{L}_z\rangle}{\Omega} = \frac{2m}{\Omega}
  \int\!d^3 r\,[r_x j_y(\rv)-r_y j_x(\rv)]\,.
\end{equation}
The factor of two is a consequence of our convention that $\jv$
denotes the current density per spin state. Using again \Eqs{deformed}
and (\ref{currentexplicit}), we can express the moment of inertia in
terms of the density profile in the corresponding spherical trap as a
simple radial integral:
\begin{multline}
\Theta = \frac{8\pi m}{3}\Big(\frac{\omegab^2}{\omega_x^2}+
  \frac{\omegab^2}{\omega_y^2}\Big)\int_0^\infty dr\,r^4
  \Big[\rhob_n(r)\\
  +\Big(\frac{\omega_x^2-\omega_y^2}{\omega_x^2+\omega_y^2}\Big)^2
  \rhob_s(r)\Big]\,.
\label{theta}
\end{multline}

\begin{figure}
\includegraphics[scale=1.4]{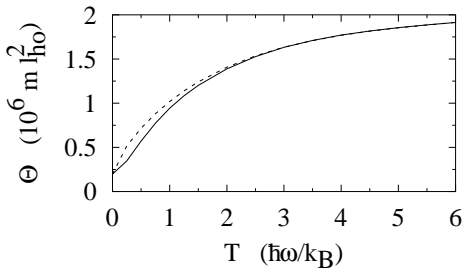}
\caption{\label{Fig4} Moment of inertia, $\Theta$ of a trapped Fermi
gas as function of temperature. The parameters are the same as in
\Figs{Fig2} and \ref{Fig3}. The moment of inertia and the temperature
are given in harmonic oscillator units, i.e., $\Theta$ in $ml_{ho}^2$
and $T$ in $\hbar\omegab/k_B$. The transition temperature to the BCS
phase lies at approximately $5.5\hbar\omegab$. The solid line has been
obtained by performing a full HFB calculation for $\rhob_0(r)$ and
$\Deltab_0(r)$, while for the dashed line the TF results have been
used.}
\end{figure}

In \Fig{Fig4} we show the moment of inertia for the same set of
parameters that were already used in \Figs{Fig2} and \ref{Fig3} as a
function of temperature. The solid line has been calculated by using
the HFB density profiles, while the dashed line was obtained from the
density profile within TF approximation. One can see that the moment
of inertia decreases strongly as the temperature goes to zero. The
limiting value at zero temperature is determined by the deformation of
the trap in the $xy$ plane,
\begin{equation}
\Theta(T=0) =
  \Big(\frac{\omega_x^2-\omega_y^2}{\omega_x^2+\omega_y^2}\Big)^2
  \Theta_\mathit{rigid}\,,
\end{equation}
where $\Theta_\mathit{rigid}$ denotes the corresponding rigid-body
moment of inertia [which can be obtained from \Eq{theta} by putting
$\rhob_n = \rhob_0$ and $\rhob_s = 0$]. In our case of
$\omega_x/\omega_y = 0.7$, we have $\Theta(T=0) \approx 0.12
\Theta_\mathit{rigid}$. An important point to notice is that, coming
from high temperatures, one does not observe an appreciable change of
the moment of inertia until one reaches temperatures far below the
critical one. The reason for this effect is that the main contribution
to the moment of inertia comes from the outer regions of the trapped
gas, where the order parameter becomes small and where consequently
the normal-fluid fraction is large even far below $T_c$. The
discrepancy between the HFB (solid line) and TF results (dashed line)
below $\approx 2 \hbar\omegab$ can be traced back to the effect that
within the TF approximation the gap near the surface vanishes already
at very low temperature, such that the normal-fluid fraction near the
surface is overestimated within TF.

To conclude, we have applied the two-fluid model known from the theory
of superconductivity \cite{Leggett,BetbederMatibet,Schrieffer} to the
case of ultracold trapped fermionic atoms in the BCS phase. In
contrast to the usual situation, the ratio of the normal and
superfluid densities is explicitly position dependent due to the
inhomogeneous trapping potential. Specializing to the case of a slowly
rotating system, we have shown that the linear order in $\hbar$ of the
linear response equations gives a current which can be decomposed in a
natural way into normal and superfluid components. The normal
component appears as a consequence of Cooper pairs which are broken by
thermal excitations already below the critical temperature $T_c$. We
have shown that especially the outer region of the trapped gas behaves
essentially as if it was normal-fluid, even far below $T_c$. Only the
central region of the gas keeps its superfluid character up to $T_c$.

As a consequence, the moment of inertia decreases more slowly as it
was previously expected \cite{UrbanSchuck} if the temperature is
lowered below $T_c$, i.e., the effects of superfluidity become visible
only far below $T_c$. This important but in a certain sense negative
result will apply analogously to other observables which are mainly
sensitive to the surface of the system, like, e.g., collective
modes. For example, the theory presented here was used in
\Ref{GrassoUrbanVinas} in order to explain the temperature dependence
of the strength of the response function for the so-called ``twist
mode''. There the effect was even more dramatic, since the relevant
integral contained an $r^6$ weight factor instead of $r^4$ in
\Eq{theta}.

We are therefore convinced that it is not justified to compare the
experimentally measured frequencies of collective modes directly with
theoretical predictions obtained for zero temperature, as it is done
in the current literature \cite{Bartenstein,Kinast}. We rather expect
that the temperature dependence is important and can be predicted by
generalizing the two-fluid model presented here to the dynamic case,
i.e., by performing the Wigner-Kirkwood expansion of the
time-dependent HFB (TDHFB) equations up to linear order in $\hbar$
\cite{BetbederMatibet,Woelfle}. This leads to a generalization of the
Vlasov equation for the normal phase, which results from the
Wigner-Kirkwood expansion of the time-dependent Hartree-Fock (TDHF)
equation up to linear order in $\hbar$.

However, one should keep in mind that the Thomas-Fermi approximation
for the ground state as well as the Wigner-Kirkwood expansion of the
dynamical equations (i.e., the generalized Vlasov equation and the
superfluid hydrodynamics to which it reduces in the zero-temperature
limit) depend on the assumption $\hbar\omega_i\ll\Delta$ for $i =
x,y,z$. Concerning the validity of the Thomas-Fermi approximation for
the ground state, the condition $\hbar\omega_i\ll\Delta$ has been
inferred from the requirement that the coherence length $\xi = \hbar
v_F/\pi\Delta$ ($v_F$ being the Fermi velocity) must be much smaller
than the typical length scale of the system, which is approximately
given by the Thomas-Fermi radius $R_{TF} = \sqrt{2\mu/m\omega_i^2}$
\cite{Heiselberg}. However, it is less evident where the assumption
$\hbar\omega_i\ll\Delta$ enters into the description of the dynamics
of the system within the generalized Vlasov equation.

To give a specific example, in \Ref{UrbanSchuck}, quantum corrections
to the moment of inertia of higher orders in $\hbar (\omega_x \pm
\omega_y) / \Delta$ were discussed. Also in the case of the strength
of the twist mode mentioned above, the fully quantum-mechanical
(``microscopic'') calculation showed deviations from the two-fluid
model, especially at very low temperatures. In both cases, the
corrections act as if the normal-fluid component of the system was
larger than predicted by \Eq{rhon} and in particular non-vanishing
even at zero temperature. In a certain sense the accelerations acting
on the Cooper pairs during their motion through the inhomogeneous
potential seem to have a similar pair-breaking effect as the thermal
excitations which are responsable for the normal-fluid component given
by \Eq{rhon}. From a completely different point of view, \Eq{rhon} is
usually derived by looking at the long-wavelength limit ($q\xi\ll 1$)
of the current-current correlation function in a homogeneous system
\cite{FetterWalecka,Schrieffer}. In the trapped system, however, the
wave vectors must be of the order $q\approx 1/R_{TF}$, and we recover
the condition $\xi\ll R_{TF}$.

Deviations from superfluid hydrodynamics ($T = 0$) or from the
two-fluid model ($T > 0$), respectively, may therefore be especially
important in the case of the strongly elongated traps used in current
experiments, which have rather high radial trapping frequencies
$\omega_x$ and $\omega_y$. Therefore this kind of quantum effects
should be studied in more detail. In the case of collective modes,
this could be done, e.g., by comparing systematically the results
obtained in quantum mechanical quasiparticle RPA (QRPA) calculations
\cite{BruunMottelson} with those of hydrodynamics.
\begin{acknowledgments}
I thank P. Schuck for fruitful discussions and critical reading of
the manuscript. I also acknowledge useful discussions with S. Sinha.
\end{acknowledgments}

\end{document}